\documentclass[]{aa}

\usepackage{natbib}
\usepackage{xcolor}
\usepackage{graphicx}
\usepackage{dcolumn}
\usepackage{bm}
\usepackage{aas_macros}
\usepackage{mathtools}
\usepackage{booktabs}
\usepackage[varg]{txfonts}

\newcommand{\reaction}[6]{\nuc{#1}{#2}(#3,#4)\/\nuc{#5}{#6}}
\newcommand{\nuc}[2]{\ensuremath{^{#1}}{#2}}

\newcommand{\Ercm}[1]{\ensuremath{E_{r}^{\text{c.m.}} = #1}~keV}
\newcommand{\capg}[0]{\reaction{39}{Ca}{p}{$\gamma$}{40}{Sc} }
\newcommand{\arpg}[0]{\reaction{35}{Ar}{p}{$\gamma$}{36}{K} }
\newcommand{\nap}[0]{\reaction{13}{N}{$\alpha$}{p}{16}{O} }

\graphicspath{{./figs/}}

\begin{document}


\title{Correlated Energy Uncertainties in Reaction Rate
  Calculations}

\author{Richard Longland\inst{1,2} \and Nicolas de S\'er\'eville\inst{3}}
\institute{North Carolina State University, Raleigh, NC 27695 \and
 Triangle Universities Nuclear Laboratory, Durham, NC
  27708 \and
Universit\'e Paris-Saclay, CNRS/IN2P3, IJCLab, 91405 Orsay, France}


\date{\today}
\abstract{Monte Carlo methods can be used to evaluate the uncertainty
  of a reaction rate that arises from many uncertain nuclear
  inputs. However, until now no attempt has been made to find the
  effect of correlated energy uncertainties in input resonance
  parameters.}{To investigate the impact of correlated energy
  uncertainties on reaction rates.}{ Using a combination of numerical
  and Monte Carlo variation of resonance energies, the effect of
  correlations are investigated. Five reactions are considered: two
  fictional, illustrative cases and three reactions whose rates are of
  current interest.}{The effect of correlations in resonance energies
  depends on the specific reaction cross section and temperatures
  considered. When several resonances contribute equally to a reaction
  rate, and are located either side of the Gamow peak, correlations
  between their energies dilute their effect on reaction rate
  uncertainties. If they are both located above or below the
  maximum of the Gamow peak, however, correlations
  between their resonance energies can increase the reaction rate
  uncertainties. This effect can be hard to predict for complex
  reactions with wide and narrow resonances contributing to the
  reaction rate.}{}


\maketitle

\section{Introduction}
\label{sec:intro}

Thermonuclear reaction rates dictate energy generation and elemental
synthesis in stars and stellar explosions. They are a key physical
input to computational stellar models that attempt to explain
astrophysical phenomena in conjunction with observational data. It is
for this reason that reaction rates must be well known, and moreover,
their uncertainties must be well understood if comparisons between
stellar models and observations are to be made reliably. Reaction rate
uncertainties were first addressed by the Nuclear Astrophysics
Compilation of REaction rates (NACRE) collaboration in an
evaluation of 86 reactions on A = 1 to 30 nuclei \citep{ANG99}. In
that time period more sophisticated techniques were developed to
quantify reaction rate uncertainties
\citep{Thompson1999,Iliadis1999}. Those techniques were used in the
reaction rate evaluation of \cite{ILI01}. While that method was
developed to accurately propagate cross section uncertainties to
nuclear reaction rates, some approximations were necessary. For
example, the propagation of energy uncertainties required that they
are ``very small''~\citep{Thompson1999}. The method was also not able
to propagate the uncertainty in rates dominated by broad resonances.

Over the last decade, advances have been made in characterising
thermonuclear reaction rate uncertainties using statistically rigorous
practices. \cite{LON10} introduced a Monte Carlo method for
propagating the uncertainties in resonant reaction cross sections to
reaction rates by carefully assigning probability density
distributions to individual experimental inputs. Similar methods have
been developed in, for example, \cite{Rauscher2012}. The method of
\cite{LON10} was further improved by \cite{SAL13} and extended to
account for uncertain spin-parities \citep{MOH14}. The methods have
been applied to nucleosynthesis in a number of environments,
e.g. \cite{LON12c,RAU16,Cescutti2018}, including Monte Carlo variations
in full hydrodynamic
models~\citep{Fields2016,Fields2018}. Correlations between input
parameters was recently addressed by \cite{Longland2017}, who
introduced a formalism for accounting for resonance parameters that
are correlated through some standard normalisation point. That study
did not, however, address correlated resonance energies. Resonance
energies can be strongly correlated, particularly for unobserved
resonances located according to excitation energies in the compound
nucleus. If the Q-value of the reaction is poorly known, those
resonance energies will also be equally unknown and strongly
correlated. This is particularly important for radioactive nuclei
where there resonances are rarely measured directly. Another common
case arises from resonance energies determined from excitation
function measurement using stable beams with moderately well-known
beam energy. If the beam energy is selected using an analysing or
switching magnet the resonance energies are highly correlated with
uncertainties dictated by the calibration of said magnet.

In this paper, we will extend the formalism laid out in
\cite{Longland2017} to account for correlations in resonance
energies. A brief overview of the reaction rate formalism will be
outlined in Sec.~\ref{sec:rate-formalism}, particularly as it pertains
to resonance energies. The Monte Carlo reaction rate uncertainty
propagation method is reviewed in Sec.~\ref{sec:MC-propagation} and
the method for computing correlated resonance energies is developed. The method is tested with illustrative
examples in Sec. \ref{sec:cases} and real reactions in
Sec.~\ref{sec:real-tests}. A summary of our findings and
recommendations is given in Sec.~\ref{sec:summary}.

\section{Reaction Rate Formalism}
\label{sec:rate-formalism}

What follows is a brief overview of astrophysical reaction
  rates. For more detail the reader is encouraged to refer to
  \cite{CAULDRONS,ANG99,LON10,SAL13,ILIBook}. The reaction rate
per particle pair, $\langle \sigma v \rangle$, is defined as
\begin{equation}
  \label{eq:reactionrate}
  \langle \sigma v \rangle = \left(\frac{8}{\pi \mu}\right)^{1/2}
  \frac{1}{(kT)^{3/2}}\int_0^{\infty} E \sigma(E) e^{-E/kT} dE,
\end{equation}
where $\mu$ is the reduced mass of the system, $k$ is the Boltzmann
constant, $T$ is the temperature at which the reaction rate is being
calculated, and $\sigma(E)$ is the energy-dependent cross section of
the reaction. 

The cross section can be parameterised by removing the s-wave Coulomb
barrier tunnelling cross section as
\begin{equation}
  \label{eq:astro-sfactor}
  \sigma(E) = \frac{1}{E} S(E) e^{-2 \pi \eta},
\end{equation}
where $\eta$ is the Sommerfeldt parameter defined by 
\begin{equation}
  \label{eq:sommer}
  2 \pi \eta = 0.98951 Z_0 Z_1 \sqrt{\frac{\mu}{E}}.
\end{equation}
$Z_0$ and $Z_1$ are the atomic numbers of the interacting
particles. The quantity $S(E)$ is the so-called Astrophysical
S-Factor, which contains any details of the cross section that are not
accounted for by simple s-wave Coulomb barrier scattering. We see from
combining Eqs. \ref{eq:astro-sfactor} and \ref{eq:reactionrate}
that the reaction rate is defined as
\begin{equation}
  \label{eq:reactionrate-sfac}
  \langle \sigma v \rangle = \left(\frac{8}{\pi \mu}\right)^{1/2}
  \frac{1}{(kT)^{3/2}}\int_0^{\infty} S(E) e^{-2 \pi \eta} e^{-E/kT} dE.
\end{equation}
The product of the two exponentials: the Gamow factor,
$e^{-2 \pi \eta}$ and the Boltzmann factor, $e^{-E/kT}$
approximates the energy range over which the astrophysical
  S-factor should be known. This product is the ``Gamow peak'', and
  the location of the maximum is at $E_0$. This maximum and the width
  of the Gamow peak are defined by:
\begin{align}
  E_0 =& \left[\left(\frac{\pi}{\hbar}\right)^2 (Z_0 Z_1 e^2)^2
         \left(\frac{\mu}{2}\right)(kT)^2\right]^{1/3} = 0.1220 \left( Z_0^2
    Z_1^2 \mu T_9^2 \right)^{1/3} \label{eq:E0}\\
  \Delta =& \frac{4}{\sqrt{3}} \sqrt{E_0 kT} = 0.2368 \left( Z_0^2
    Z_1^2 \mu T_9^5 \right)^{1/6}, \label{eq:Delta}
\end{align}
where $Z_0$ and $Z_1$ are the atomic numbers of the interacting
nuclei, $e$ is the elementary electric charge, and $T_9$ is the
temperature in $10^9$ K.

Many reactions of astrophysical importance proceed through nuclear
resonances, populating compound nuclear states that subsequently
decay. Note that we only consider 2-body reactions in the
following. For a single, isolated resonance, the cross section in
Eq.~(\ref{eq:reactionrate}) can be replaced by
\begin{equation}
  \label{eq:resonance-xsec}
  \sigma(E) = \frac{2 J + 1}{(2 J_1 + 1)(2 J_2 + 1)} \frac{\pi}{k^2}\frac{\Gamma_a(E) \Gamma_b(E)}{(E-E_r)^2 + \Gamma^2/4}.
\end{equation}

Here, $J$, $J_1$, and $J_2$ are the spins of the resonance, target,
and projectile particles, respectively. This angular momentum term is
often denoted by the symbol, $\omega$.  $\Gamma_a(E)$ and
$\Gamma_b(E)$ are energy-dependent quantities describing the entrance
and exit partial widths of the state in question. For example, for a
(p,$\gamma$) reaction, $\Gamma_a(E)$ corresponds to the proton partial
width and $\Gamma_b(E)$ is the $\gamma$-ray partial width. $\Gamma$
corresponds to the \textit{total} width of the state, and $E_r$ is the
resonance energy. For a charged particle, $\Gamma_a(E)$ is defined as
\begin{equation}
  \label{eq:Gammap}
  \Gamma_a(E) = 2\frac{\hbar^2}{\mu R^2} P_{\ell}(E)\, C^2S\, \theta_{a}^2,
\end{equation}
where $P_{\ell}(E)$ is the energy-dependent penetration factor
describing the probability of the particles tunnelling though
  the Coulomb barrier, $C^2S$
is the product of the Isospin Clebsh-Gordan coefficient
for the interacting particles and spectroscopic
factor. This latter quantity describes how well an excited
  state in the compound nucleus can be described by a single-particle state.
$\theta_a^2$ is the dimensionless single-particle reduced width, which
can be calculated theoretically from the particle's
  wave-function (see \cite{ILI97}). We consider this latter quantity to
be unity, here, since we're only considering relative effects under
energy variations. The channel radius, $R$ is calculated as
$R = R_0 \left( A_p^{1/3} + A_t^{1/3} \right)$. This choice
  does not have a large effect on the calculations \textit{provided}
  it is used consistently throughout.

If the resonance is sufficiently narrow such that its partial widths
do not vary significantly over its width, we can assume that
$\Gamma_a$ and $\Gamma_b$ are constant. In that case, the integral in
Eqn.~\ref{eq:reactionrate} can be evaluated algebraically. Now the
resonance cross section is replaced by a single quantity: the
resonance strength, $\omega \gamma$, which is defined by
\begin{equation}
  \label{eq:omegagamma}
  \omega \gamma = \omega \frac{\Gamma_p \Gamma_{\gamma}}{\Gamma_p + \Gamma_{\gamma}}.
\end{equation}
It's important to recognise that the partial widths, $\Gamma_p$ and
$\Gamma_{\gamma}$, in Eqn. \ref{eq:omegagamma} depend on resonance
energies (through Eqn. \ref{eq:Gammap} for charged particles). Any
energy shifts must be carefully propagated through these equations to
determine shifted resonance strengths. Once this procedure is
followed, Eq.~(\ref{eq:reactionrate}) can be replaced with
\begin{equation}
  \label{eq:reactionrate-narrowresonance}
  \langle \sigma v \rangle = \left(\frac{2\pi}{\mu kT}\right)^{3/2}
  \hbar^2 \sum_i \omega \gamma_i e^{-E_r/kT}
\end{equation}
This paper focuses on reaction rates dominated by resonances in the
absence of interference. Addressing interfering resonance reaction
rates that are best described by R-matrix or other complex models is
left for future work.

\section{Monte Carlo Reaction Rates}
\label{sec:MC-propagation}

\subsection{Formalism}
\label{sec:mc-formalism}

In order to investigate the influence that correlated resonance energy
uncertainties have on reaction rates, the Monte Carlo reaction rate
method first described in~\cite{LON10} was used as a starting
point. The general strategy laid out in that study was to first assign
a probability density distribution to every uncertain input parameter
to the reaction rate calculation. For the case of resonant cross
sections, these uncertain parameters are the resonance energies,
$E_r$, partial widths, $\Gamma_a$, and resonance strengths,
$\omega \gamma$, in Eqns.~\ref{eq:resonance-xsec},
\ref{eq:omegagamma},
and~\ref{eq:reactionrate-narrowresonance}. The strategy was extended
to allow for uncertain spin-parities in ~\cite{MOH14}.

Once probability density distributions have been obtained for
uncertain input parameters, sample parameters are randomly chosen from
the distributions assuming that all parameters are independent (the
validity of this assumption is discussed below). The
reaction rate calculated from the sampled parameters represents a
single sample rate. This procedure is repeated many times (10,000 is
preferred but at least 3000 samples was found to produce stable
results) to obtain a distribution of reaction rates that can be
summarised with a reaction rate probability density
distribution. \cite{LON10} found that the reaction rate
probability density can often be summarised with a log-normal
distribution with shape parameters $\mu$ and $\sigma$. The recommended
rate is then given by:
\begin{equation}
  \label{eq:recommendedrate}
  \langle \sigma v \rangle_{\text{rec.}} = e^{\mu(T)}.
\end{equation}
The ``low'' and ``high'' rates given by the 1-$\sigma$ uncertainties
are found using
\begin{equation}
  \label{eq:highlowrate}
  \langle \sigma v \rangle_l = e^{\mu(T)}e^{-\sigma(T)} \qquad \langle \sigma v \rangle_h = e^{\mu(T)}e^{+\sigma(T)}
\end{equation}

The procedure described above was found to be a flexible method for
estimating the reaction rate uncertainties. Provided energy
uncertainties are correctly propagated into partial width
uncertainties, it works equally well for narrow and wide resonant
reaction rates with large uncertainties. However, the method did not
account for the case in which there are correlations between
parameters.

\subsection{Correlated Energies}
\label{sec:correlations}

Recently, efforts were made to include the effects of correlated
partial width and resonance strength uncertainties in the Monte Carlo
method described above~\citep{Longland2017}. This case of correlated
widths and resonance strengths needed to be investigated because
resonances are usually normalised to some standard, well known
resonance. Given that experimental reality, the assumption of
independent variables made in ~\cite{LON10} is not strictly
accurate. 

Consider the following example: a partial width for resonance $j$,
$\Gamma_j$, is correlated with a reference resonance, $r$. These
partial widths carry factor uncertainties,
$f.u._{j} \equiv \sigma_{\Gamma,j}/\Gamma_j$ and
$f.u._{r} \equiv \sigma_{\Gamma,r}/\Gamma_r$, respectively where $\sigma_{\Gamma,j}$ is
the uncertainty in $\Gamma_j$, etc.  A single correlation parameter,
$\rho_j$, can be used to describe the magnitude of their correlation:
\begin{equation}
  \label{eq:rho}
  \rho_j = \frac{\sigma_{\Gamma,r}}{\Gamma_r} \frac{\Gamma_j}{\sigma_{\Gamma,j}} \equiv \frac{f.u._r}{f.u._j}.
\end{equation}

During the Monte Carlo procedure, the following steps are
  taken: First, random, uncorrelated samples are produced for the
  reference resonance and resonance $j$. These samples, denoted $x_i$
  and $y_{j,i}$, are drawn from standard normal distributions (i.e.,
  with a mean of zero and standard deviation of
  1). \cite{Longland2017} showed that a simple, 2-step procedure can
  then be used to calculate correlated partial widths. Second, the
  samples are correlated using
\begin{equation}
  y_{j,i}' = \rho_j x_{i} + \sqrt{1-\rho_j^2} y_{j,i}, \label{eq:yp}
\end{equation}
where $y_{j,i}'$ are the correlated samples for resonance
  $j$. Finally, the partial width samples can be computed with the
  knowledge that partial widths should be log-normally distributed
  (see \cite{LON10}):
\begin{equation}
  \Gamma_{j,i} = \Gamma_{j,\text{rec.}} \, (f.u.)^{y'_{j,i}}. \label{eq:widthsamples}
\end{equation}
Here, the recommended partial width for resonance $j$ is
$\Gamma_{j,\text{rec.}}$.

Note that the probability density distributions defined by
Eq.~\ref{eq:widthsamples} are a log-normal distribution. For small
uncertainties they resemble Gaussian distributions, but are not
defined for negative values. They therefore describe physical
parameters such as partial widths and resonance strengths
appropriately. They are also well motivated by the central limit
theorem, as discussed in detail in ~\cite{LON10}.

In the present study we follow the same strategy as above, but must
consider two modifications: (i) energy uncertainties must be
propagated through partial widths. For the charged particle reactions
considered here, this is accomplished using Eq.~\ref{eq:Gammap}; and
(ii) energy uncertainties are not log-normal. Often large resonance
energy uncertainties arise because of large uncertainties in reaction
Q-values. For example, imagine a low-energy resonance
with a large energy uncertainty that has an appreciable probability of
being a sub-threshold resonance. Modification (i) is already accounted
for in the \texttt{RatesMC} code implementation of the Monte Carlo
methods outlined in ~\cite{LON10} \footnote{Note that the correlation between resonance energy and partial widths makes reaction rate uncertainty propagation using calculated resonance strengths unreliable. Partial widths should be used whenever available.} The second case requires 
modifications to Eqns.~\ref{eq:rho} and~\ref{eq:widthsamples}:
\begin{align}
  \rho_j =& \frac{\min(\sigma_E)}{\sigma_{E,j}}.   \label{eq:rho-energy}\\
  E_{j,i} =& E_{j,\text{rec.}} + y_{j,i}' \sigma_{E,j}. \label{eq:correlated-E-sample}
\end{align}
The correlated standard normal samples, $y_{j,i}'$ are computed using
Eq.~\ref{eq:yp}. The purpose of Eq.~\ref{eq:rho-energy} is similar
to that of Eq.~\ref{eq:rho}. It ensures that correlations between
resonance energies do not exceed their experimental limits. For
example, consider a reaction in which the majority of the resonance
energy uncertainty arises from an uncertain reaction Q-value,
$\sigma_Q$.
Those resonance energies will be correlated with
$\rho \approx 1$. Now assume another resonance in that reaction
populates an excited state that \textit{also} has an uncertain
excitation energy, $\sigma_{Ex}$. The energy of that resonance will
have a larger overall uncertainty (given by
$\sigma^2 = \sigma_Q^2 + \sigma_{Ex}^2$) and will not be fully
correlated with the other resonance energies. For this example
resonance, $\rho < 1$.

Finally, cases in which resonance energies are determined through
different means must be considered. In most cases, high energy
resonance energies are determined through direct measurement of the
reaction cross section. Resonance energies can be determined by
measuring a yield curve, for example. Low-energy resonances in the
same reaction may be determined from excitation energies and an
uncertain Q-value. Therefore, the ability to enable or disable energy
correlations for a resonance must be available. The \texttt{RatesMC}
code has been modified to allow this.

\section{Test Cases}
\label{sec:cases}

To investigate the effect of correlated energy uncertainties on
reaction rates, two fictional
reactions designed to probe low- and high-density resonance
regimes are first investigated. The effect on actual reaction rates will be detailed in Sec.~\ref{sec:real-tests}.

\subsection{Low Resonance Density}
\label{sec:cases-low-density}

\begin{figure*}[ht]
  \centering
  \includegraphics[width=0.8\textwidth]{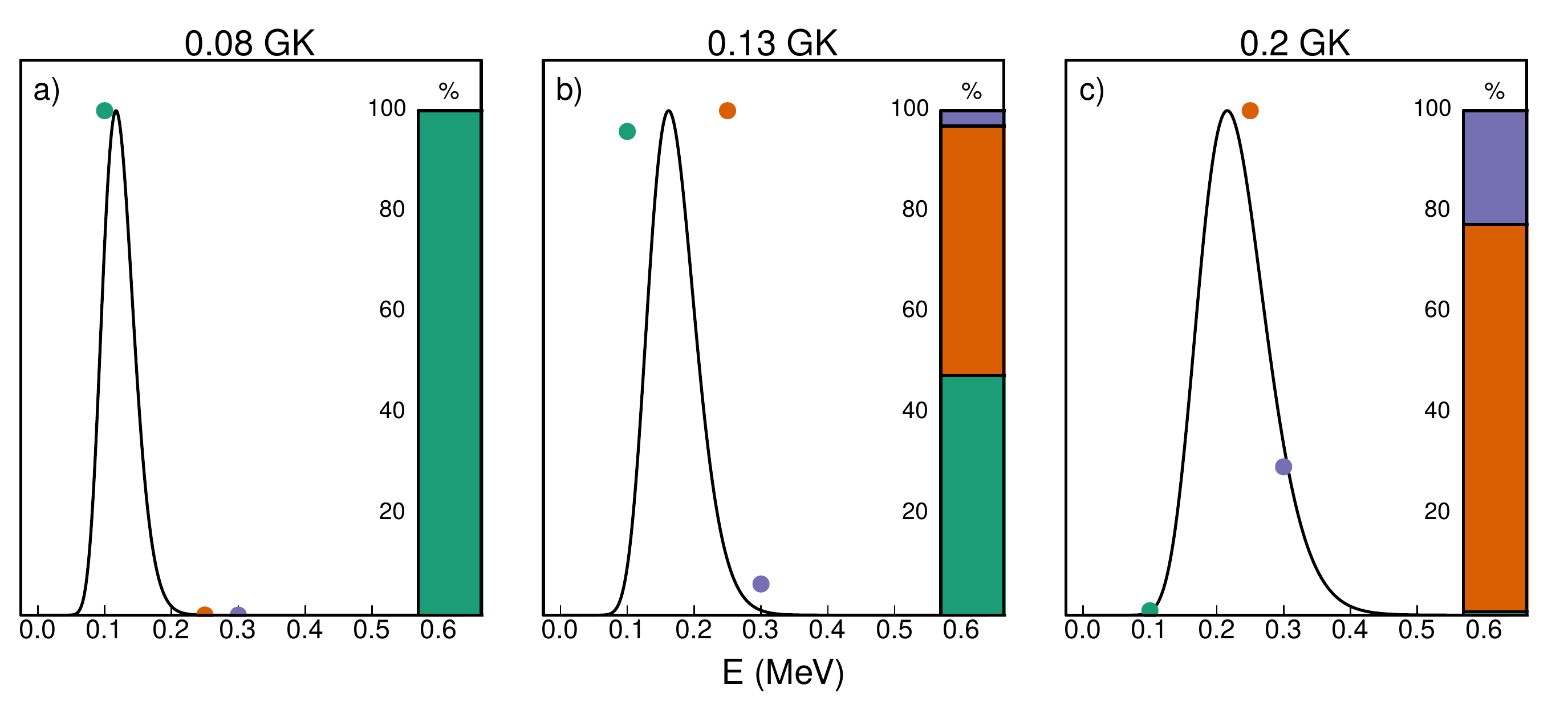}
  \caption{(colour online) Test cases for our investigation of the
    impact of energy uncertainty correlations on nuclear reaction rate
    uncertainties. Shown as a solid line is the Gamow peak
      in arbitrary units calculated from
    Eqn. \ref{eq:reactionrate-narrowresonance} at three
    temperatures: 0.08 GK, 0.13 GK, and 0.2 GK in panels a), b), and
    (c) respectively. The reaction rate for each of the three
    narrow resonances described in Tab.~\ref{tab:test-resonances} are
    shown as coloured points. The y-direction is scaled arbitrarily for
    clarity to highlight \textit{which} resonances contribute most to
    the reaction rate and where they are located in comparison to the
    Gamow peak. This information is also displayed by the bar on the
    right of each panel. For example, in panel a), only one resonance
    -- the \Ercm{100} resonance shown in green -- contributes to the
    reaction rate. In panel b), two resonances -- \Ercm{100} and
    \Ercm{250} in green and orange -- contribute approximately 50\%
    each to the reaction rate at 0.13 GK.}
  \label{fig:3Temps}
\end{figure*}

The first test case considered is shown in Fig. \ref{fig:3Temps}. The
physical system consists of a fictional reaction whose cross section
consists entirely of three isolated, narrow resonances. Thus,
Eqn.~\ref{eq:reactionrate-narrowresonance} can be used to calculate
the reaction rate. The partial widths are calculated using
Eqn.~\ref{eq:Gammap} by assuming R$_0 = 1.25$~fm and a (p,$\gamma$)
reaction occurring on an isotope with an atomic number of 12 and mass
of 23. The resonance parameters chosen are shown in
Tab.~\ref{tab:test-resonances}.

\begin{table}[h]
  \centering
  \begin{tabular}{lccccc}
    \hline \hline
    $E_r^{\text{c.m.}}$ (keV) & J$^{\pi}$ & C$^2$S & $\Gamma_p$ (eV)      & $\Gamma_{\gamma}$ (eV) & $\omega \gamma$ (eV)  \\ \hline
    100.0        & 1$^-$     & 1      & $2.8\times 10^{-7}$ & 3.0                    & $1.0\times 10^{-7}$ \\
    250.0        & 1$^-$     & 1      & $2.0\times 10^{-1}$  & 3.0                    & $7.0\times 10^{-2}$ \\
    300.0        & 1$^-$     & 1      & $1.5\times 10^{0}$  & 3.0                    & $3.8\times 10^{-1}$  \\
    \hline \hline
  \end{tabular}
  \caption{Parameters of the three resonances considered in our
    fictional test case. $E_r^{\text{c.m.}}$ is the centre-of-mass
    resonance energy (in keV).}
  \label{tab:test-resonances}
\end{table}

These resonances contribute different amounts to the reaction rate
depending on the temperature. The low-energy charged-particle reaction
resonances most important at low temperatures can be well predicted by
the Gamow peak defined in
Eqn.~\ref{eq:reactionrate-sfac}. Indeed, this is the case
as shown in Fig.~\ref{fig:3Temps}, where the three resonance
contributions to the total reaction rate follow closely the
progression of the Gamow peak as temperatures increase. Higher energy
resonances, though, will not follow this pattern. Their cross
sections become constrained by $\gamma$-ray partial widths, which does
not exhibit the Coulomb barrier energy
dependence~\citep{Newton2007}. 

\begin{figure*}[ht]
  \centering
  \includegraphics[width=0.7\textwidth]{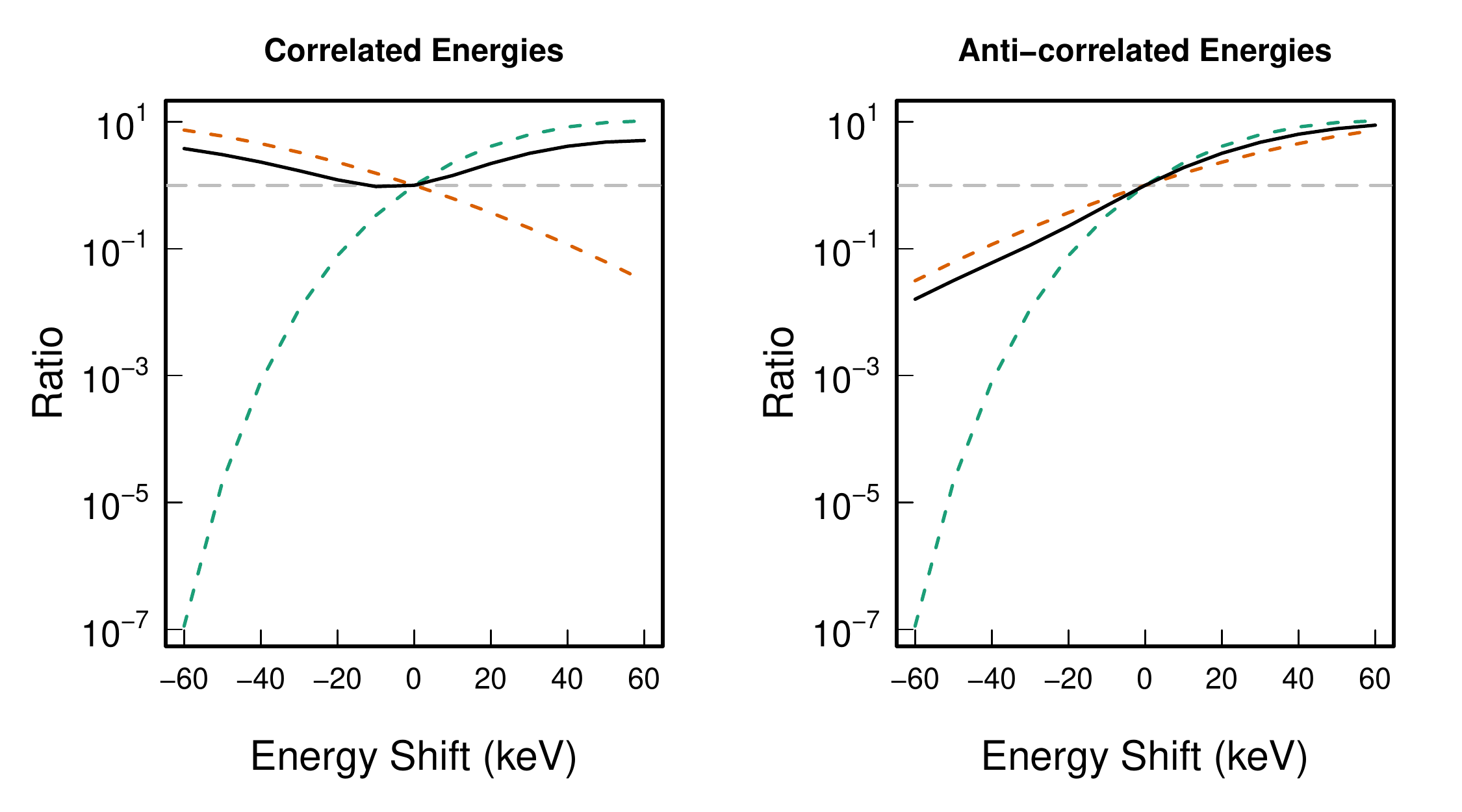}
  \caption{(colour online) The effects of correlated and
    anti-correlated energy variations to a pair of resonances on the
    calculated reaction rate at 0.13 GK. The green and orange dashed
    lines show the individual resonance contributions from the
    \Ercm{100} and \Ercm{250} resonances to the total reaction rate
    shown by the black line. The Energy shift is defined as the shift
    of the lower energy resonance. All are normalised to their
    recommended values at $\Delta E = 0$.}
  \label{fig:EnergyVary}
\end{figure*}

First, consider the reaction rate at $T=0.13$~GK. At this temperature,
the two resonances at \Ercm{100} and \Ercm{250} contribute
approximately equally to the total reaction rate as shown by the bar
on the right of the centre panel. They are situated either side of the
maximum of the Gamow peak. To investigate the effect that
their energy uncertainties have on the reaction rate, their resonance
energies are varied over a given range. For each trial resonance
energy, the proton partial width is re-calculated using
Eqn.~\ref{eq:Gammap} and the parameters in
Tab.~\ref{tab:test-resonances}. Using this, the reaction rate is
determined using Eqs.~\ref{eq:omegagamma}
and~\ref{eq:reactionrate-narrowresonance}.  These resonance energies
are varied using two schemes: (i) correlated energies, so any increase
in the energy of one resonance corresponds to an equal increase in the
other, and (ii) anti-correlated energies, in which any energy
\textit{increase} in one resonance energy corresponds to an equal
magnitude \textit{decrease} in the other. The resonance energies are
varied by $\pm 60$~keV this way. The variations affect the individual
contributions to the rate as well as the total rate. These are shown
in Fig.~\ref{fig:EnergyVary}.

Figure~\ref{fig:EnergyVary} shows that larger reaction rate
variations are expected if resonance energies are
anti-correlated. To understand this effect, consider first the
correlated energy case in the left-hand panel as well as the middle
panel of Fig.~\ref{fig:3Temps}. As the resonances both increase in energy,
the one at \Ercm{100} shifts closer to the maximum of the Gamow peak,
thus increasing its contribution to the total rate. Conversely, the
resonance at \Ercm{250} moves \textit{away} from the maximum of the Gamow peak and
\textit{decreases} its contribution. The net effect is that the total
rate does increase, but the magnitude is weakened by the opposite
contributions of the two resonances. As the resonance energies
decrease, a similar effect is apparent: the \Ercm{100} resonance
contributes less while the \Ercm{250} resonance contributes more,
resulting again in an increase in reaction rate that is weakened by
the opposite contributions.

In the case of anti-correlated resonance energies in the right-hand
panel of Fig.~\ref{fig:EnergyVary}, the resonances contributions work
in tandem. When the \Ercm{100} resonance energy is decreased, it moves
away from the maximum of the Gamow peak to lower energies, while the \Ercm{250}
resonance \textit{also} moves away, but to higher energies. Thus the
contributions of both resonances decrease, resulting in a reduced
total reaction rate. Similarly, as one moves towards the maximum of
the Gamow peak,
so will the other, resulting in a strengthened increase in the total
rate. Anti-correlated energy uncertainties in this case result in an
increased reaction rate uncertainty. 

\begin{figure}[ht]
  \centering
  \includegraphics[width=0.4\textwidth]{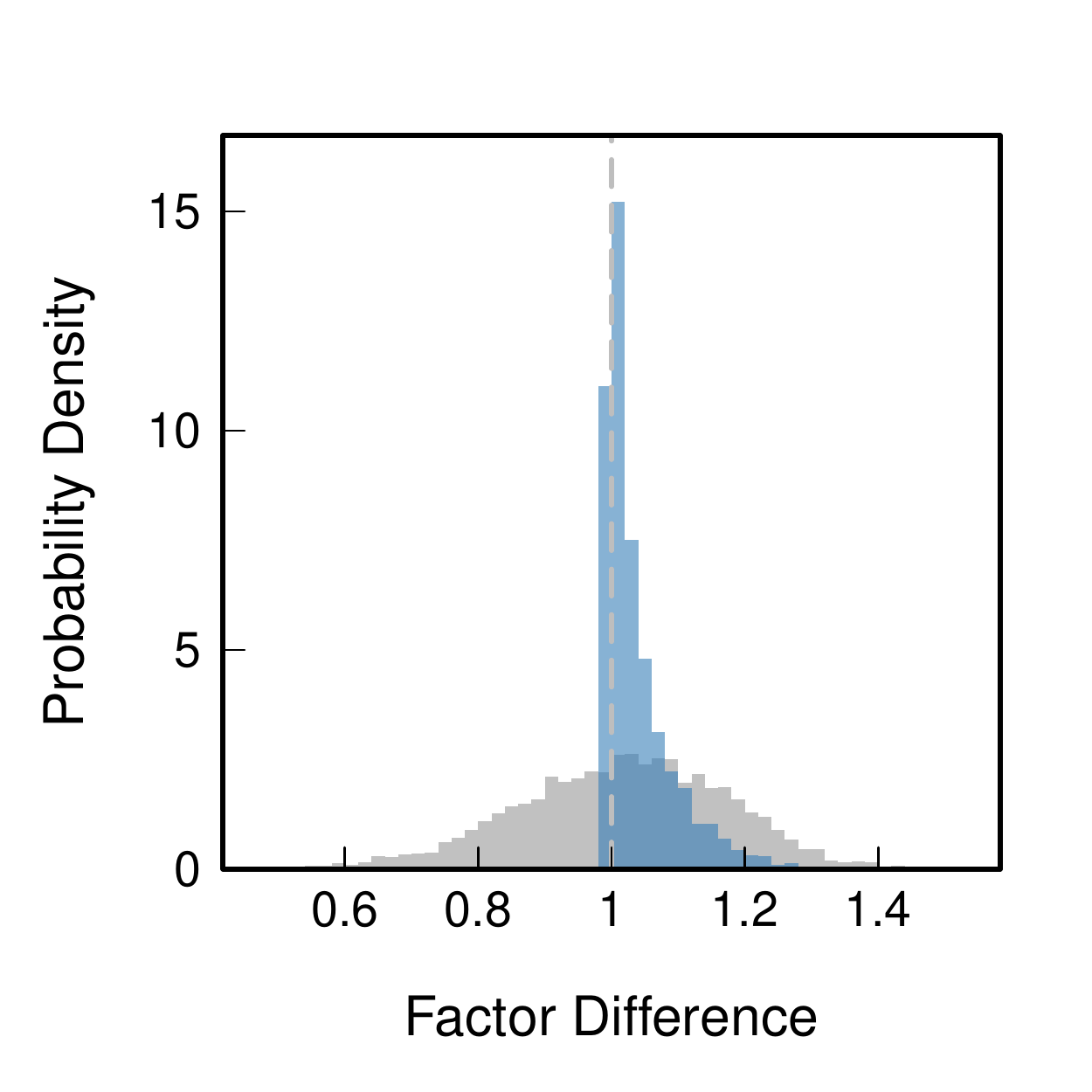}
  \caption{(colour online) Reaction rate probability distributions from
    the Monte Carlo variation of resonance energies at 0.13 GK. Shown
    in grey is the uncorrelated case, in which the resonance energies
    are allowed to vary independently. In blue is the case in which
    the resonance energies are fully correlated. The distribution
    becomes narrower and highly skewed in this case owing to the
    effect illustrated in Fig.~\ref{fig:EnergyVary}.}
  \label{fig:MCVary}
\end{figure}

The anti-correlated resonance energy example discussed above will
rarely occur in experimental resonance energy measurements. However,
if the resonance energies are treated as completely uncorrelated
during the Monte Carlo procedure, their relative variations will be
somewhere between the fully correlated and fully anti-correlated
cases. Thus, taking into account the effect illustrated in
Fig.~\ref{fig:EnergyVary} we expect larger uncertainties for
uncorrelated resonance energies than correlated energies. This is
indeed the case, as shown in Fig.~\ref{fig:MCVary}, which shows the
reaction rate probability density distributions for these two
cases. In grey, a broad, approximately Gaussian peak represents the
probability density distribution for the uncorrelated energy case. In
the correlated case, the effect shown in Fig.~\ref{fig:EnergyVary} is
clear in that the probability distribution is not only narrower, but
is also highly skewed. Since we chose a temperature at which the
resonances are either side of the Gamow peak and contribute
approximately equally to the total rate, the reaction rate can
\textit{only} increase as their energies are varied.

\begin{figure*}[ht]
  \centering
  \includegraphics[width=0.7\textwidth]{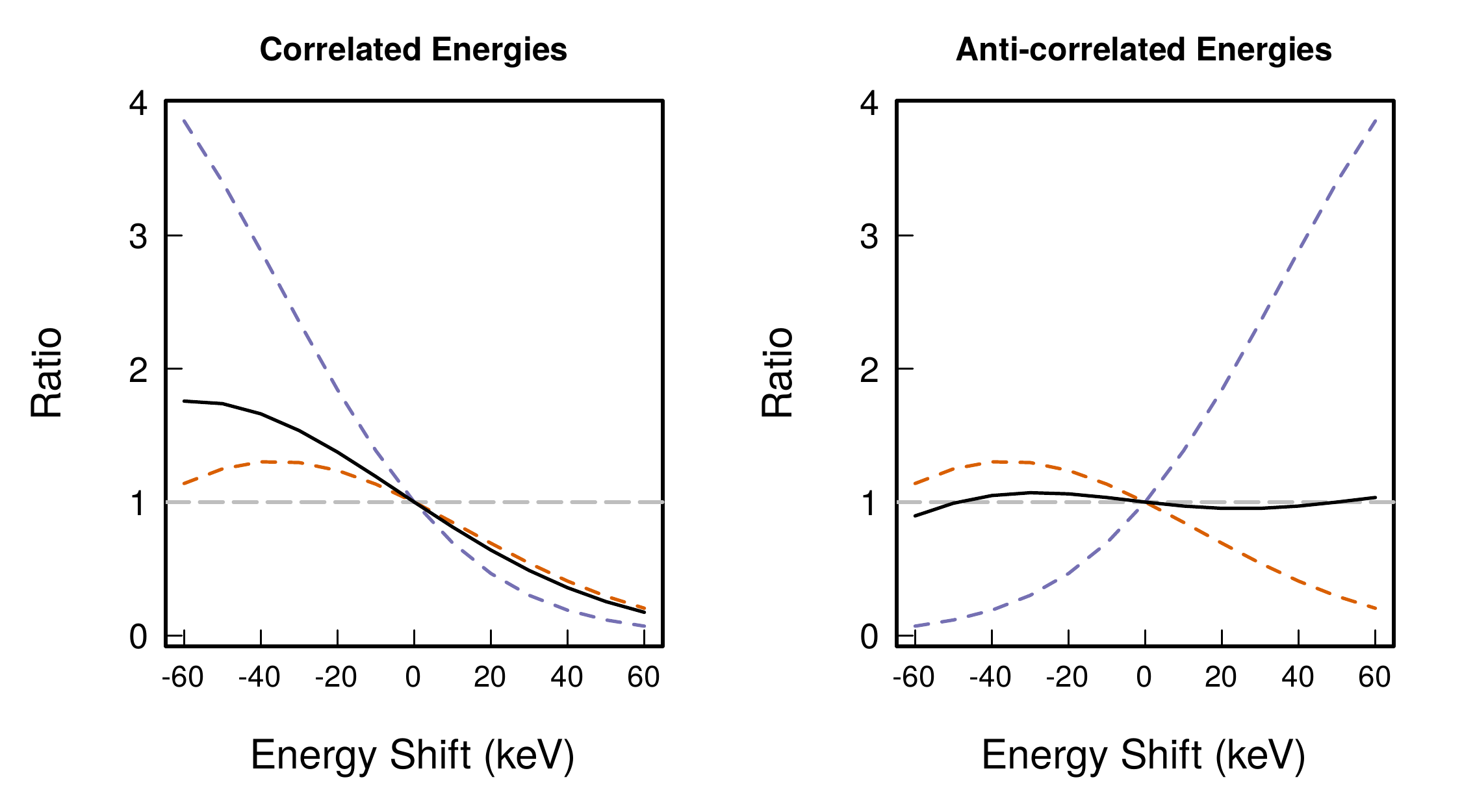}
  \caption{(colour online) The effects of correlated and
    anti-correlated resonance energies on the reaction rate at
    $T=0.2$~GK. The red and blue dashed lines correspond to the
    contributions from the \Ercm{250} and \Ercm{300} resonances,
    respectively. See Fig.~\ref{fig:EnergyVary} for details.}
  \label{fig:EnergyVary-highT}
\end{figure*}

How universal is this effect? In the example above, a specific
temperature was chosen to correspond to two resonances either side of
the Gamow peak. To investigate more possibilities, the
third panel in Fig.~\ref{fig:3Temps} is illustrative. In this case,
the resonances at \Ercm{250} and \Ercm{300} contribute about 75\% and
25\% to the reaction rate, respectively. They're also both located
above the maximum of the Gamow peak, so as we shift their
energies in a correlated manner, they will both shift toward or away
from the peak in unison. The effect of their variations on the total
reaction rate is shown in Fig.~\ref{fig:EnergyVary-highT}. In this
case, the opposite effect to that observed in
Fig.~\ref{fig:EnergyVary} is apparent. If the resonances are
correlated, they move together to reduce or increase the reaction
rate. If they are anti-correlated, their contributions essentially
cancel out to produce very little variation in the total reaction
rate.

\begin{figure}[ht]
  \centering
  \includegraphics[width=0.4\textwidth]{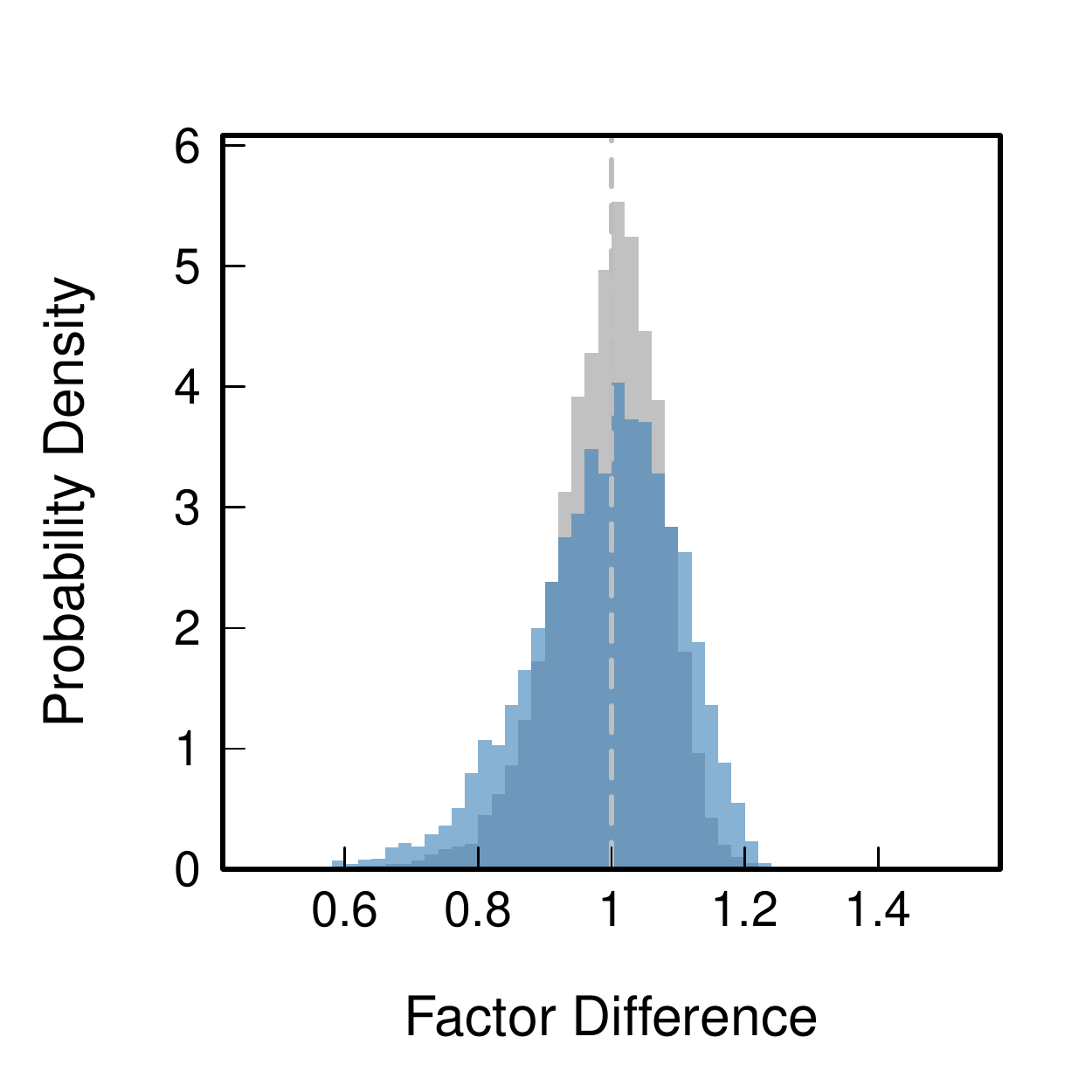}
  \caption{(colour online) Reaction rate probability distributions from
    the Monte Carlo variation of resonance energies at 0.2 GK. See
    Fig.~\ref{fig:MCVary} for details. The probability distribution
    becomes wider when correlated energies are considered in this case
    owing to the effect illustrated in
    Fig.~\ref{fig:EnergyVary-highT}.}
  \label{fig:MCVary-highT}
\end{figure}

The Monte Carlo reaction rate comparison at $T=0.2$~GK is shown in
Fig.~\ref{fig:MCVary-highT}. In this case, we see the opposite effect
to the example at $T=0.13$~GK. If energy correlations are taken into
account, the reaction rate uncertainties \textit{increase}. Clearly,
these effects are hard to predict, particularly when large resonance
energies are concerned. However, using Monte Carlo uncertainty
propagation, we are able to account for the effect of energy
correlations on the reaction rate uncertainties.

\subsection{High Resonance Density}
\label{sec:cases-high-density}

Following the same procedure for a high density of resonances produces
results that are easier to predict. In this case, the reaction rate
uncertainty when correlated energy uncertainties are taken into
account reliably \textit{decreases}. Now, resonance are placed at
energies between \Ercm{50} and \Ercm{400} with a spacing of
20~keV. 

\begin{figure}[ht]
  \centering
  \includegraphics[width=0.4\textwidth]{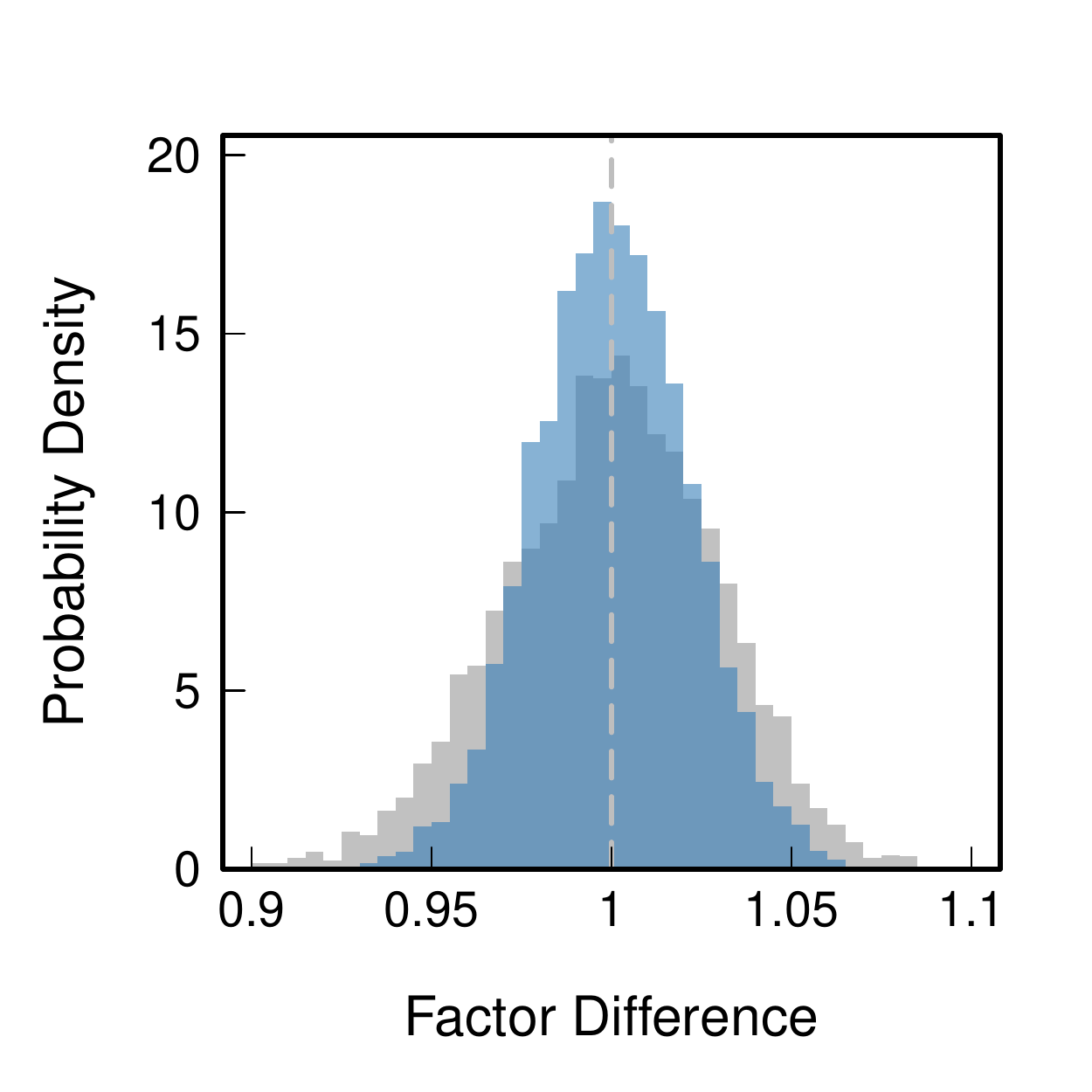}
  \caption{(colour online) Reaction rate probability distributions from
    the Monte Carlo variation of resonance energies at 0.13 GK for the
    high resonance density case. See Fig. \ref{fig:MCVary} for
    details.}
  \label{fig:MCVary-manyRes}
\end{figure}

The reaction rate uncertainty is shown in
Fig. \ref{fig:MCVary-manyRes} for $T = 0.13$~GK. 
The uncertainty for correlated energies (blue) clearly decreases in
comparison to the uncorrelated case (grey). As temperature increases
we find that the effect of correlations decreases because more
resonances contribute to the rate.

\subsection{Discussion}
\label{sec:discussion}

The results discussed above are difficult to predict a
priori. However, a conservative estimate of reaction rate
uncertainties would only be concerned with the case in which resonance
energy correlations \textit{increase} the reaction rate
uncertainties. This case is shown in
Fig. \ref{fig:MCVary-highT}. For these cases to occur, the
  resonances should be located on the same side of the Gamow peak
  (i.e., as their energy increases/decreases, correlations would cause
  them all to increase/decrease their contributions to the total rate
  in unison). This requires more than one resonance contributing to
  the rate, but the resonance density cannot be too high, else
  resonances would be distributed throughout the Gamow peak. For
  example, it would not be realistic to expect a case where a high
  resonance density is found on one side of the Gamow peak, but no
  resonances on the other. Furthermore, the resonance energy
  uncertainty in contributing resonances should be large in comparison
  to the Gamow peak's width, defined by Eq.~(\ref{eq:Delta}).

\section{Physical Cases}
\label{sec:real-tests}

Now that the general behaviour of how reaction rate uncertainties
change when resonance energy correlations are taken into account, some
physically realistic cases will be considered. These are the
\reaction{35}{Ar}{p}{$\gamma$}{36}{K} reaction; the
\reaction{39}{Ca}{p}{$\gamma$}{40}{Sc} reaction; and the
\reaction{13}{N}{$\alpha$}{p}{16}{O} reaction. They span a range of
resonance densities and represent cases where the resonance energies
are known to be uncertain and correlated.

\subsection{The \reaction{35}{Ar}{p}{$\gamma$}{36}{K} Reaction}

The \reaction{35}{Ar}{p}{$\gamma$}{36}{K} reaction is a key
  reaction in explosive hydrogen burning. In x-ray bursts this
  reaction is expected to occur faster than its competing
  $\beta$-decay, but in novae (i.e., lower temperatures) the rate is
  less well known. Its effect on the nucleosynthesis of heavier
  elements is not well understood~\citep{Glasner2009}. The reaction
rate was evaluated in \cite{Iliadis1999}. At that time, the Q-value of
this reaction was poorly known \citep{Audi1995}, leading to large,
correlated resonance energy uncertainties. Since that time the
excitation energy uncertainties in \nuc{36}{K} have been dramatically
reduced by an order of magnitude by \cite{Wrede2010}. However, the
resonance energies are still expected to be correlated and the
resonance density of this reaction is very low with just 4 known
resonances below 1 MeV. For these reasons it is an ideal case with
which to investigate correlated resonance energies in the Monte Carlo
framework. The resonance parameters from \cite{Wrede2010} are listed
in Tab.~\ref{tab:35Ar-pg-resonances}. Note that we have assigned very
small uncertainties (1\%) to the partial widths so that the resonance
energy effects can be clearly identified. Separate
  calculations confirm that the uncertainties due to other sources sum
  quadratically, as expected.

\begin{table}[h]
  \centering
  \begin{tabular}{lccc}
    \toprule \toprule
    $E_r^{\text{c.m.}}$ (keV) & J$^{\pi}$ & $\Gamma_p$ (eV)      & $\Gamma_{\gamma}$ (eV)   \\ \midrule
    48.4 (8)       & 2$^-$     & $3.2\times 10^{-109}$ & $2.7 \times 10^{-4}$  \\
    259.9 (9)     & 2$^+$     & $5.7\times 10^{-7}$   & $1.0 \times 10^{-2}$  \\
    538.5 (9)     & 3$^-$     & $4.2\times 10^{-1}$   & $4.7 \times 10^{-4}$  \\
    623.4 (9)     & 2$^+$     & $2.5\times 10^{0}$    & $6.8 \times 10^{-3}$  \\
    \bottomrule \bottomrule
  \end{tabular}
  \caption{Resonance parameters for the
    \reaction{35}{Ar}{p}{$\gamma$}{36}{K} reaction taken from
    \cite{Wrede2010}. The uncertainties in $\Gamma_p$ and
    $\Gamma_{\gamma}$ have been assumed to be 1\% to emphasise the
    effect of the energy uncertainties (see text).}
  \label{tab:35Ar-pg-resonances}
\end{table}

\begin{figure}[ht]
  \includegraphics[width=0.48\textwidth]{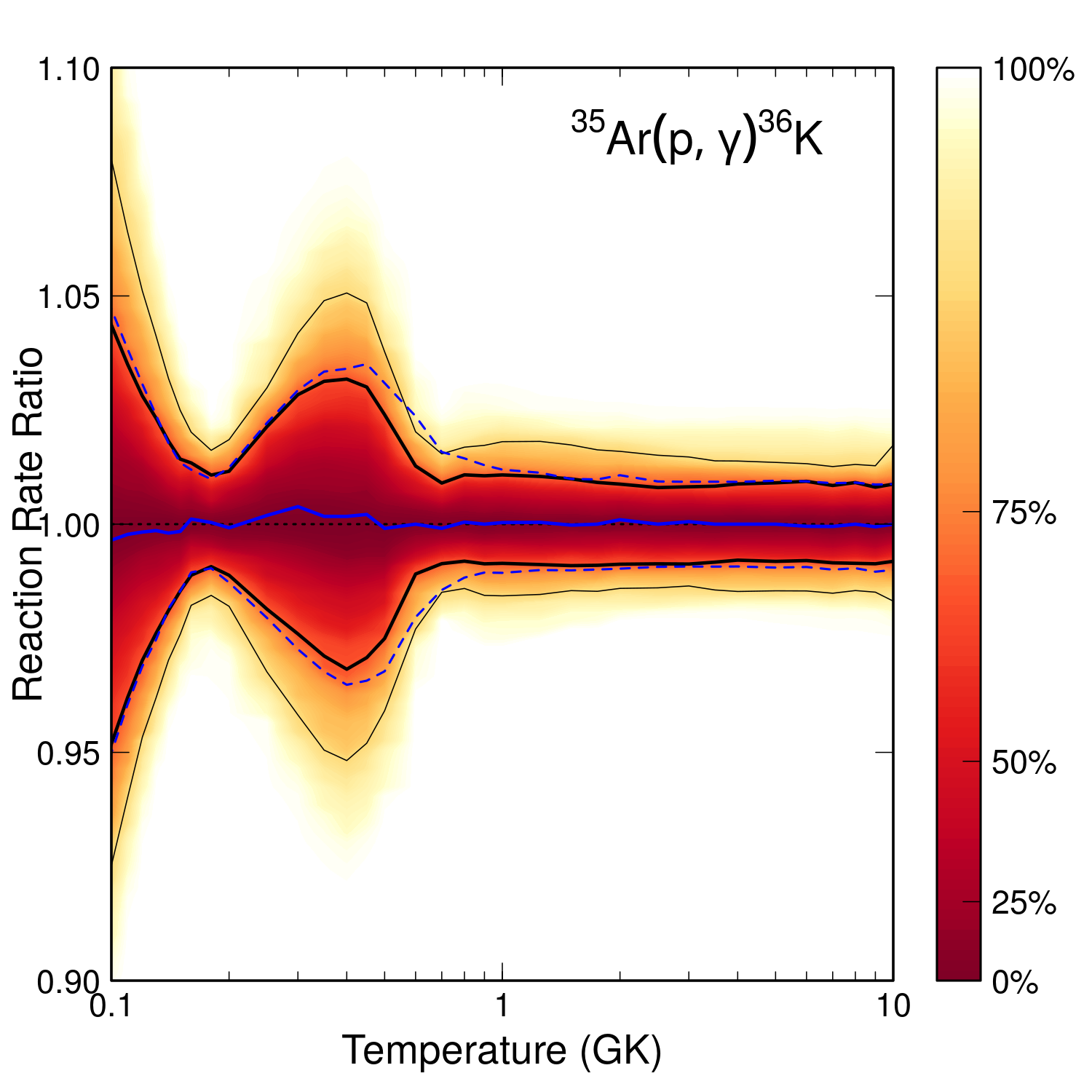}
  \caption{\label{fig:GraphContour-35Arpg}(colour online) Reaction rate
    uncertainties for the \reaction{35}{Ar}{p}{$\gamma$}{36}{K}
    reaction assuming the resonance parameters shown in
    Tab.~\ref{tab:35Ar-pg-resonances}. Recall that \textit{only}
    resonance energies are taken into account. The rate has been
    normalised to the median rate, which is shown by a dashed line at
    unity. The thick and thin black lines represent the $1\sigma$ and
    $2\sigma$ uncertainties in the correlated resonance energy
    calculation. The blue lines show the uncorrelated case, with the
    thick line representing the median rate (again, normalised to the
    recommended correlated energy rate), and dashed lines showing the
    $1\sigma$ uncertainties. At $T=400$~MK, the effect of correlations
    slightly decreases the reaction rate uncertainty.}
\end{figure}

The reaction rate uncertainties for the
\reaction{35}{Ar}{p}{$\gamma$}{36}{K} reaction assuming the resonance
parameters shown in Tab.~\ref{tab:35Ar-pg-resonances} are shown in
Fig. \ref{fig:GraphContour-35Arpg}. The coloured contour represents the
reaction rate uncertainties arising from correlated energy
uncertainties, with thick and thin lines representing the $1\sigma$
and $2\sigma$ uncertainty bands, respectively. These rates have been
normalised to the recommended (median) rate, which is shown by a
horizontal line at unity. The blue lines show the reaction rate
uncertainties when resonance energy correlations are \textit{not}
taken into account. The thick blue line represents the median rate,
and the thin dashed blue lines represent the $1\sigma$ uncertainties.
This figure shows that over most of the temperature range, energy
correlations do not strongly affect the reaction rate
uncertainties. They are only slightly smaller when taking resonance
energy correlations into account. This is mostly due to the fact that
below 200 MK and above 1 GK, only one resonance is contributing and
the effect of energy correlation is almost in-existent. In between
these temperatures two resonances contribute to the reaction rate. At
400 MK, for example, the resonances are located either side of the
maximum of the Gamow peak. In this case the effect of
correlations is small, but in line with the case described in
Sec. \ref{sec:cases-low-density}: as the \Ercm{260} resonance moves to
a lower energy, for example, the \Ercm{623} resonance also moves to a
lower energy, thus reducing the impact of resonance energies on the
reaction rate.

These calculations were also performed assuming the (obsolete)
resonance parameters reported in \cite{Iliadis1999}. Additionally the
total rate uncertainty is much larger owing to the larger energy
uncertainties, the correlations between those energies have the same,
minor, effect on the uncertainty as outlined above. 

\subsection{The \reaction{39}{Ca}{p}{$\gamma$}{40}{Sc} Reaction}
\label{sec:cases-39Ca-pg}

The \capg reaction is also important in explosive
  nucleosynthesis. It influences the end-point of the rp-process in
  x-ray bursts, and has also been evaluated in
\cite{Iliadis1999}. The resonance density is similar to the \arpg
reaction, but in this case the Q-value is better known. The resonance
energy uncertainties are just 5-6 keV, as shown in
Tab.~\ref{tab:39Ca-pg-resonances}.

\begin{table}[h]
  \centering
  \begin{tabular}{lccc}
    \toprule \toprule
    $E_r^{\text{c.m.}}$ (keV) & J$^{\pi}$ & $\Gamma_p$ (eV)      & $\Gamma_{\gamma}$ (eV)   \\ \midrule
    223 (5)     & 2$^-$     & $2.0\times 10^{-9}$   & $1.6 \times 10^{-3}$  \\
    353 (5)     & 5$^-$     & $3.0\times 10^{-7}$   & $5.3 \times 10^{-4}$  \\
    1128 (6)    & 2$^-$     & $2.2\times 10^{2}$    & $1.3 \times 10^{-3}$  \\
    1128 (6)    & 1$^-$     & $3.0\times 10^{2}$    & $8.8 \times 10^{-4}$  \\
    \bottomrule \bottomrule
  \end{tabular}
  \caption{Resonance parameters for the \capg reaction taken from
    \cite{Iliadis1999}. Note that the two resonances at \Ercm{1128}
    correspond to a triplet structure observed in the
    \reaction{40}{Ca}{\nuc{3}{He}}{t}{40}{Sc} measurement by
    \cite{Schulz1971}. The uncertainties in $\Gamma_p$ and
    $\Gamma_{\gamma}$ have been assumed to be 1\% to emphasise the
    effect of the energy uncertainties (see text).}
  \label{tab:39Ca-pg-resonances}
\end{table}

In this case, the predictions in Sec.~\ref{sec:discussion} indicate
that there should, indeed, be an effect of resonance energy
correlations on the reaction rate. The average resonance energy
separation is 300 keV compared with $\Delta = 200$~keV at 400 MK. In
contrast to the \arpg reaction, though, there is a temperature range
at which both resonances at \Ercm{223} and \Ercm{353} lay on the
low-energy side of the Gamow peak. Thus we expect an effect of
correlated energies on the reaction rate. 

\begin{figure}[ht]
  \includegraphics[width=0.48\textwidth]{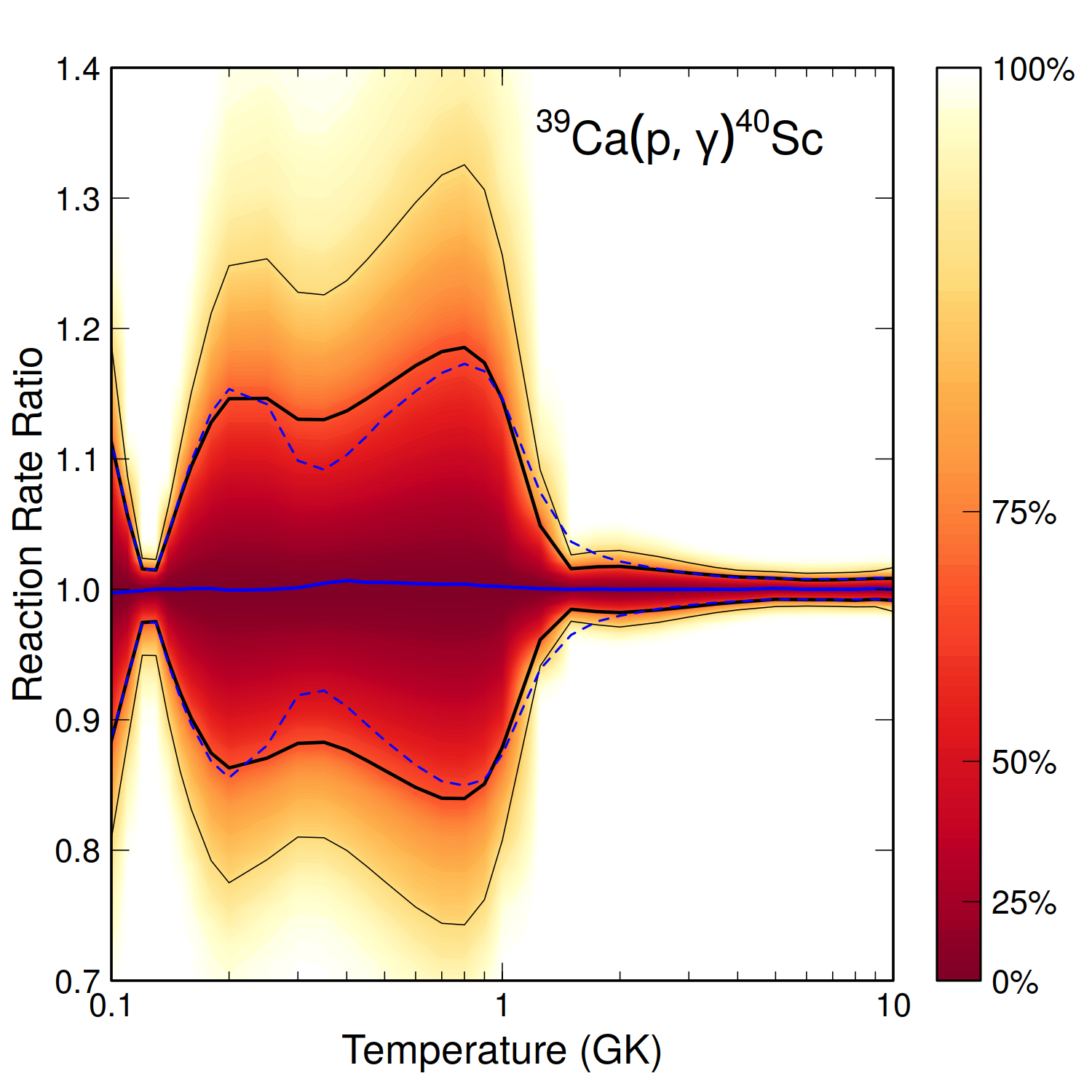}
  \caption{\label{fig:GraphContour-39Arpg}(colour online) Reaction rate
    uncertainties for the \capg reaction assuming the resonance
    parameters shown in Tab.~\ref{tab:39Ca-pg-resonances} and
    accounting \textit{only} for uncertainties in the resonance
    energies. See Fig.~\ref{fig:GraphContour-35Arpg} for description.}
\end{figure}

Figure \ref{fig:GraphContour-39Arpg} shows that correlations
between resonance energy uncertainties do, indeed, affect the reaction
rate uncertainty strongly at 300-400 MK where both resonances at
\Ercm{233} and \Ercm{353} contribute to the reaction rate and are both
are on the low-energy side of the Gamow peak. In this
particular scenario, as the resonance co-move to lower energies,
\textit{both} contribute to a lower reaction rate. Conversely if they
both co-move to higher resonance energy they both contribute to a
higher reaction rate. In the uncorrelated energy case, this scenario
is more rare, thus the rate uncertainty is smaller. Note that over
most of the temperature range, $0.2 < T_9 < 1$, the rate uncertainties
are considerable. At those temperatures, only 1 or 2 resonances ever
contributes towards the rate, and the uncertainty is arising from the
strong Coulomb barrier energy dependence in Eq.~(\ref{eq:Gammap}).

\subsection{The \reaction{13}{N}{$\alpha$}{p}{16}{O} Reaction}
\label{sec:cases-13N-ap}

The \nap reaction affects nitrogen production in supernova
  explosions as the shock-wave passes through the outer regions of the
  exploding star~\citep{Pignatari2013}. That material can eventually go
  on to form pre-solar grains, whose isotopic nitrogen ratios provide
  a precise test of astrophysical models~\citep{Zinner2014}. The \nap
  reaction rate should be known, therefore, to a high precision. The
rate was recently evaluated in~\cite{Meyer2020}. In this case the
\nuc{13}{N}+$\alpha$ threshold
($S_{\alpha+\nuc{13}{\mathrm{N}}}=5817.4~(4)$~keV) is accurately known
from \cite{Tilley1993}. However the energy of the resonances in the
compound nucleus \nuc{17}{F} suffers from systematic uncertainty of
several tens of keV which introduces a strong correlation between
resonance energies. This uncertainty originates from a possible error
in the calibration of one of the magnets used during the measurement
of the excitation functions of the \reaction{16}{O}{p}{p}{16}{O}
reaction by \cite{Sal62a,Sal62b}, and the
\reaction{16}{O}{p}{p'}{16}{O} and
\reaction{16}{O}{p}{$\alpha$}{13}{N} reactions by \cite{Dan64}.
The properties of the most influential resonances with $\alpha$-particle 
partial width determined either from direct measurements or mirror symmetry 
considerations are taken from \cite{Meyer2020} and summarised in Tab.~\ref{tab:13N-ap-resonances}.

\begin{table}[h]
  \centering
  \begin{tabular}{lccc}
    \toprule \toprule
    $E_r^{\text{c.m.}}$ (keV) & J$^{\pi}$ & $\Gamma_\alpha$ (keV) & $\Gamma_p$ (keV)   \\ \midrule
    741 (20)  &  1/2$^+$  &  1.79$\times10^{-3}$   & 200  \\
    1213 (20) &  5/2$^-$  &  4.09$\times10^{-2}$   & 3.76 \\
    1664 (20) &  3/2$^+$  &  4.08                  & 790 \\
    1732 (20) &  7/2$^-$  &  1.35$\times10^{-2}$   & 30 \\
    1935 (40) &  1/2$^+$  &  11                    & 135 \\
    2255 (30) &  5/2$^+$  &  14                    &  79 \\
    2405 (40) &  3/2$^-$  &  25                    & 636 \\
    \bottomrule \bottomrule
  \end{tabular}
  \caption{Resonance parameters for the 
    \reaction{13}{N}{$\alpha$}{p}{16}{O} reaction taken from
    \cite{Meyer2020}. The uncertainties in $\Gamma_p$ and
    $\Gamma_{\gamma}$ have been assumed to be 1\% to emphasise the
    effect of the energy uncertainties (see text)}
  \label{tab:13N-ap-resonances}
\end{table}

In order to emphasise the effect of the (un)correlated uncertainties
on the resonance energy a very small uncertainty (1\%) has been
assigned to the partial widths and the tentative spin and parity have
been considered as firmed assignment. In this particular case,
  the resonances have very large (factor of 2.5) uncertainties, which
  completely dominates the energy uncertainties under investigation
  here.

\begin{figure}[ht]
  \includegraphics[width=0.48\textwidth]{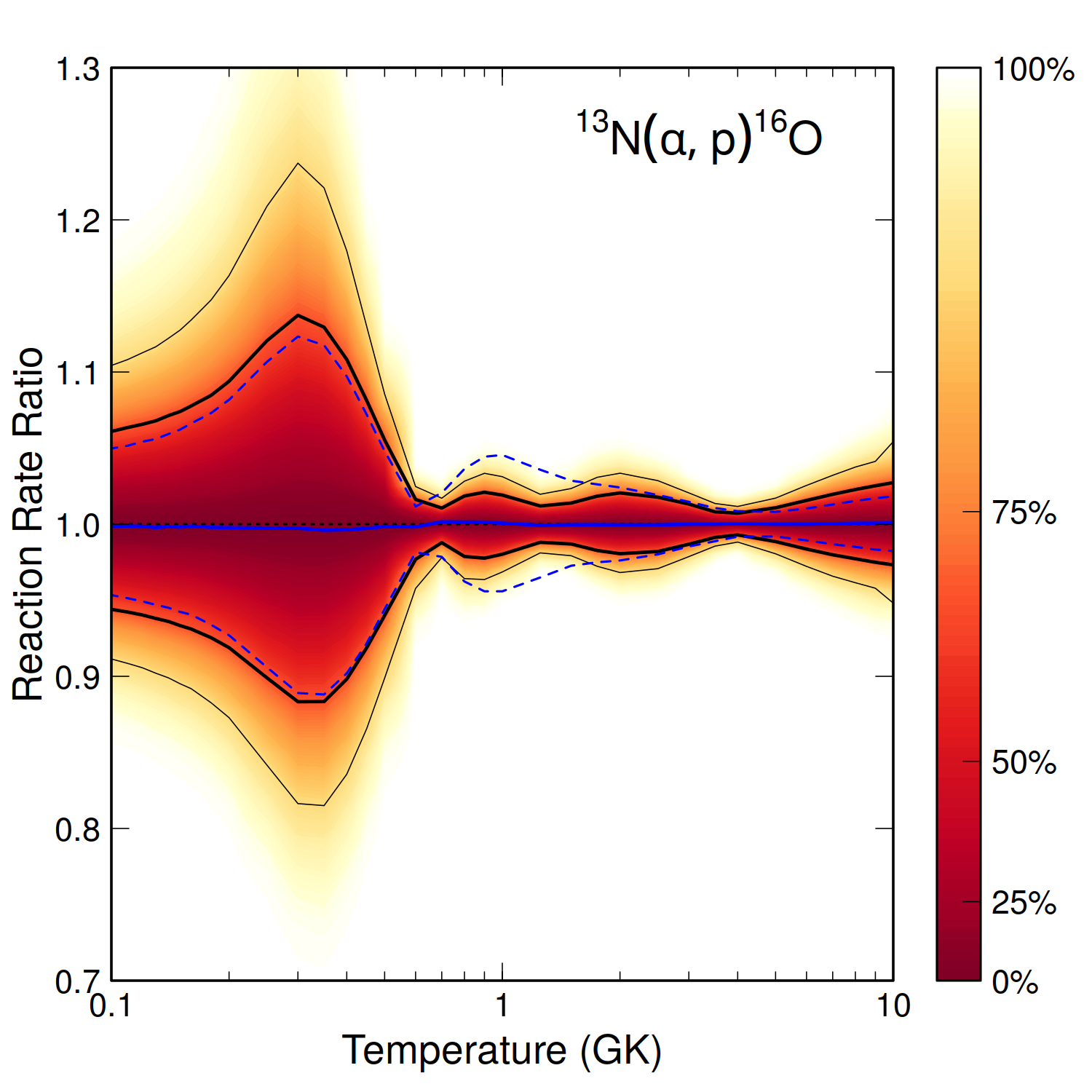}
  \caption{\label{fig:GraphContour-13Nap}(colour online) Reaction rate
    uncertainties for the \nap reaction assuming the resonance
    parameters shown in Tab.~\ref{tab:13N-ap-resonances} and
    accounting \textit{only} for uncertainties in the resonance
    energies. See Fig.~\ref{fig:GraphContour-35Arpg} for description.}
\end{figure}

Even though some resonances have large total widths their number is
relatively small and they can be considered as isolated. The case of 
low  resonance density discussed in Sec.~\ref{sec:cases-low-density} 
should then apply and an effect of correlated energies on the reaction
rate is then expected. This is indeed the case as shown in 
Fig.~\ref{fig:GraphContour-13Nap} where the \nap reaction rate
uncertainties are presented.

Let's first consider a typical temperature of 1 GK which corresponds
to a Gamow peak with a maximum of about 1~MeV and
width of $\Delta \approx 700$~keV from Eq.\
\ref{eq:Delta}. In this case the reaction rate is dominated by the two
resonances at \Ercm{741} and \Ercm{1213} which are siting on each side
of the Gamow peak's maximum. As these two resonances co-move when their energy
uncertainties are correlated, one resonance will have an increased
contribution to the reaction rate while the other one's contribution
will decrease.  A smaller reaction rate uncertainty is therefore
observed for the correlated case (black solid line in
Fig.~\ref{fig:GraphContour-13Nap}) with respect to the uncorrelated
case (dashed blue line in Fig.~\ref{fig:GraphContour-13Nap}) between
0.8~GK and 2~GK, which is in line with the findings presented in
Fig~\ref{fig:MCVary}.

The opposite behaviour is observed for temperatures lower than 400~MK
and greater than 6~GK where the uncorrelated case gives smaller
reaction rate uncertainties than in the correlated case. For these
temperatures all resonances reported in
Tab.~\ref{tab:13N-ap-resonances} are on one side of the Gamow
  peak, e.g. at higher energies for 400~MK and lower energies for
6~GK. The reaction rate therefore spans a larger range in the
correlated case inducing a greater rate uncertainty than in the
uncorrelated case as presented in
Fig.~\ref{fig:MCVary-highT}. At temperatures below about 600
  MK, the rate uncertainties are large. This is because at these
  temperatures, the \Ercm{741} resonance dominates the rate and is
  located on the low-energy tail of the Gamow peak. Any variation
  in the resonance energy has a large effect on the rate through
  Eq.~(\ref{eq:Gammap}).

\section{Summary and Conclusions}
\label{sec:summary}

Monte Carlo methods can be a powerful tool for computing statistically
rigorous uncertainties of thermonuclear reaction rates. While the
methods have been in use for some time, no effort had previously been
made to account for correlations between resonance energies. These
effects are particularly important for radioactive nuclei where
resonances are not often directly measured.

In this paper, we expanded on the correlation scheme developed in
\cite{Longland2017} to allow for correlations between resonance
energies. We found that the effects are not necessarily easy to
predict. Reactions rates dominated by many resonances are not strongly
affected by correlations, whereas those dominated by only a single
resonance at astrophysically important temperatures are also not
significantly affected. The cases that matter most are those where
multiple resonances contributed to the reaction rate. This effect is
enhanced if they are on the same side of the Gamow peak's
  maximum value.

Correlations between resonance parameters can be an important effect in
thermonuclear reaction rate calculations. The correlation of resonance
energies was previously unexplored, which has now been accounted for
in this work. Since the effects of these correlations are rather unpredictable, we recommend that any reaction rate uncertainty calculation be carefully checked to ensure corrections due to resonance energy correlations do not significantly affect the results. This will be of particular importance for reactions on isotopes far from stability, where the energies of excited states can carry large, correlated uncertainties because they are determined from uncertain reaction Q-values.

\begin{acknowledgements} 
  This material is based partly upon work supported by the
  U.S. Department of Energy, Office of Science, Office of Nuclear
  Physics, under Award Number DE-SC0017799. 
\end{acknowledgements}

\bibliographystyle{aat}
\bibliography{CorrelatedEnergies}

\end{document}